\documentclass[pra,aps,superscriptaddress,showpacs,preprint]{revtex4}%
\usepackage{amsmath}
\usepackage{amssymb}
\usepackage{amsfonts}
\usepackage{bm}
\usepackage{amssymb}
\usepackage[dvips]{graphicx}
\usepackage{dcolumn}
\usepackage{txfonts}
\usepackage{hyperref}
\usepackage{makeidx}
\usepackage{color}
\usepackage{mathtools}
\usepackage{mathrsfs}
\usepackage{subfigure}

\begin{document}

\title{Bistability and squeezing of the librational mode of an optically trapped nanoparticle}
\author{Ke-Wen Xiao}
\affiliation{Beijing Computational Science Research Center, Beijing 100193, China}

\author{Nan Zhao}
\affiliation{Beijing Computational Science Research Center, Beijing 100193, China}

\author{Zhang-qi Yin}\email{yinzhangqi@mail.tsinghua.edu.cn}
\affiliation{Center for Quantum Information, Institute for Interdisciplinary Information Sciences, Tsinghua University, Beijing 100084, China}

\begin{abstract}
We systematically investigate the bistable behavior and squeezing property of the librational mode of a levitated nonspherical nanoparticle trapped by laser beams. By expanding the librational potential to the forth order of the librational angle $\theta$, we find that the nonlinear coefficient of this mode is dependent only on the size and material of nanoparticle, but independent of trapping potential shape. The bistability and hysteresis are displayed when the driving frequency is red-detuned to the librational mode. In the blue-detuned region, we have studied squeezing of the variance of librational mode in detail, which has potential application for measurement of angle and angular momentum.

\end{abstract}

\maketitle

\section{Introduction}
Enormous research progresses of quantum optomechanics have been reported in the past few years~\cite{AspelmeyerRmp2014,KippenbergOe2007}. The applications in metrology, such as ultra-sensitive detectors for force sensing~\cite{RanjitPra2015,RanjitPra2016}, millicharge searching~\cite{MoorePrl2014} and others~\cite{RiderPrl2016,VolpePrl2006,PedaciNphys2010}, have been studied. When the motion is cooled to the quantum ground state, it can be utilized for the detection of quantum gravity~\cite{AlbrechtPra2014} and the test of objective collapse models \cite{Bassi03}. There are many different optomechanical systems, such as microtoroid~\cite{ZhangOe2013}, near-field coupled nanomechanical oscillators~\cite{AnetsbergerNphys2009}, and optical microsphere resonator~\cite{GorodetskyJosab1999}. At the same time, an optically levitated nanoparticle in vacuum begins to show its potential as a novel optomechanical system~\cite{Romero-IsartNjp2010,ChangPnas2010,LiNphys2011,LiScience2010}, which has ultra-high quality factor $Q>10^9$~\cite{LiNphys2011,JainPrl2016}. Therefore, it can be applied for ultra-sensitive measurement~\cite{YinIjmpb2013,LiScience2010}, such as the measurement of force \cite{VolpePrl2006}, torque \cite{VolpePre2007}, and mass \cite{Zhao14}, and testing the boundary between quantum and classical mechanics~\cite{Romero-IsartPrl2011,YinPra2013,Romero-IsartPra2011,Yin2017}.

A levitated nanoparticle has three translational modes and three rotational modes.
Many theoretical and experimental progress have been reported, such as the measurement of instaneous velocity of a brownian particle~\cite{LiScience2010}, the cooling of translational mode of nanoparticle~\cite{ChangPnas2010,LiNphys2011,MillenPrl2015,GieselerPrl2012}, and so on. Meanwhile, the librational(torsional) mode attracts more and more attentions.  However, these theoretical investigations were based on utilizing multiple Laguerre-Gaussian cavity modes~\cite{BhattacharyaPrl2007,Romero-IsartNjp2010,Zhou2017Optical}, or microwindmills \cite{ShiJmo2013}.
In Ref.~\cite{HoangPrl2016,Kuhn17}, the directions of the optically levitated dielectric nanoparticles were fixed, and the librational(torsional) optomechanics of a Levitated Nonspherical Nanoparticle was reported. In these experiments, the torsional mode of an ellipsoidal nanoparticle levitated by a linearly polarized Gaussian beam was experimentally observed, and the scheme of sideband cooling torsional mode was proposed \cite{HoangPrl2016}. Inspired by these experiments, researchers studied the decoherence mechanism of optically trapped nanaoparticles liberational modes \cite{Stickler16a,Zhong2016}, coupling librational modes with the internal spins \cite{Ma2016,Delord2017,nan2008cooling}, etc.

The nonlinearity of optomechanical systems offers rich new physics in both classical and quantum regimes~\cite{GieselerNphys2013,GieselerPrl2013,BrawleyNc2016,Lv2015,LiaoPra2013,ZhuPra2016,GrimsmoJop2013}, and many investigation and application of nonlinear optomechanical systems have been reported, such as ultrasensitivity optical sensor~\cite{FanOe2015}, cooling by utilising nonlinearity~\cite{FonsecaPrl2016}, and observation of bistability in a macroscopic mechanic resonator from a single chemical bond~\cite{HuangNc2016}. The nonlinearity of translational mode is usually very small, therefore could be neglected. In this paper, we systematically investigate the nonlinearity of torsional mode of a levitated nonspherical nanoparticle. We find that the nonlinearity is independent of the amplitude and frequency of the trapping laser, but only depends on the inertia of the nanoparticle. The nonlinearity could be very large when the size of the particle is small. The system can show bistability by driving the librational mode. It's worth mentioning that the red-detuning driving laser causes bistability of torsional mode, and blue-detuning laser dose not. By carefully tuning the driving detuning, the librational mode could be prepared into the squeezed state, and be used for the precision measurements of the angle and the angular momentum.

This paper is organized as follows: In Sec.\ref{SectII}, we  theoretically investigate the nonlinear effect of librational mode of nonspheric nanoparticle trapped by laser beams and deduce its Hamiltonian. In Sec.\ref{SectIII}, we study the steady-state of this nonlinear system and find librational mode have bistability. In Sec.\ref{SectIV}, we derive the linearized Hamiltonian and have studied the squeezing property of this system. In the last section Sec.\ref{SectV}, we give a brief conclusion and perspective.

\section{Model}
\label{SectII}
\begin{figure}
\centering
\subfigure{\includegraphics[height=2.2in,width=2.5in]{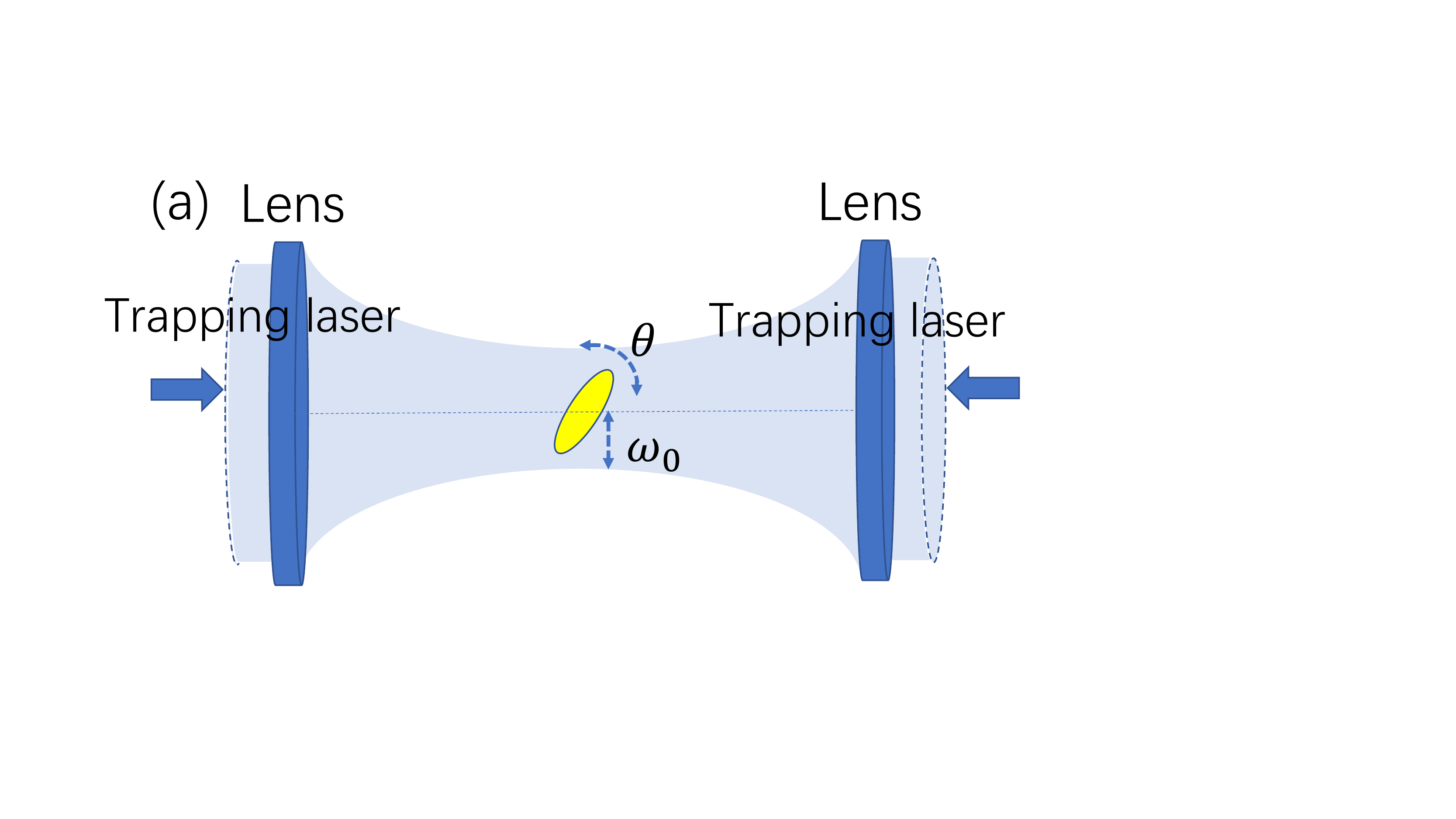}}
\subfigure{\includegraphics[width=0.48\textwidth]{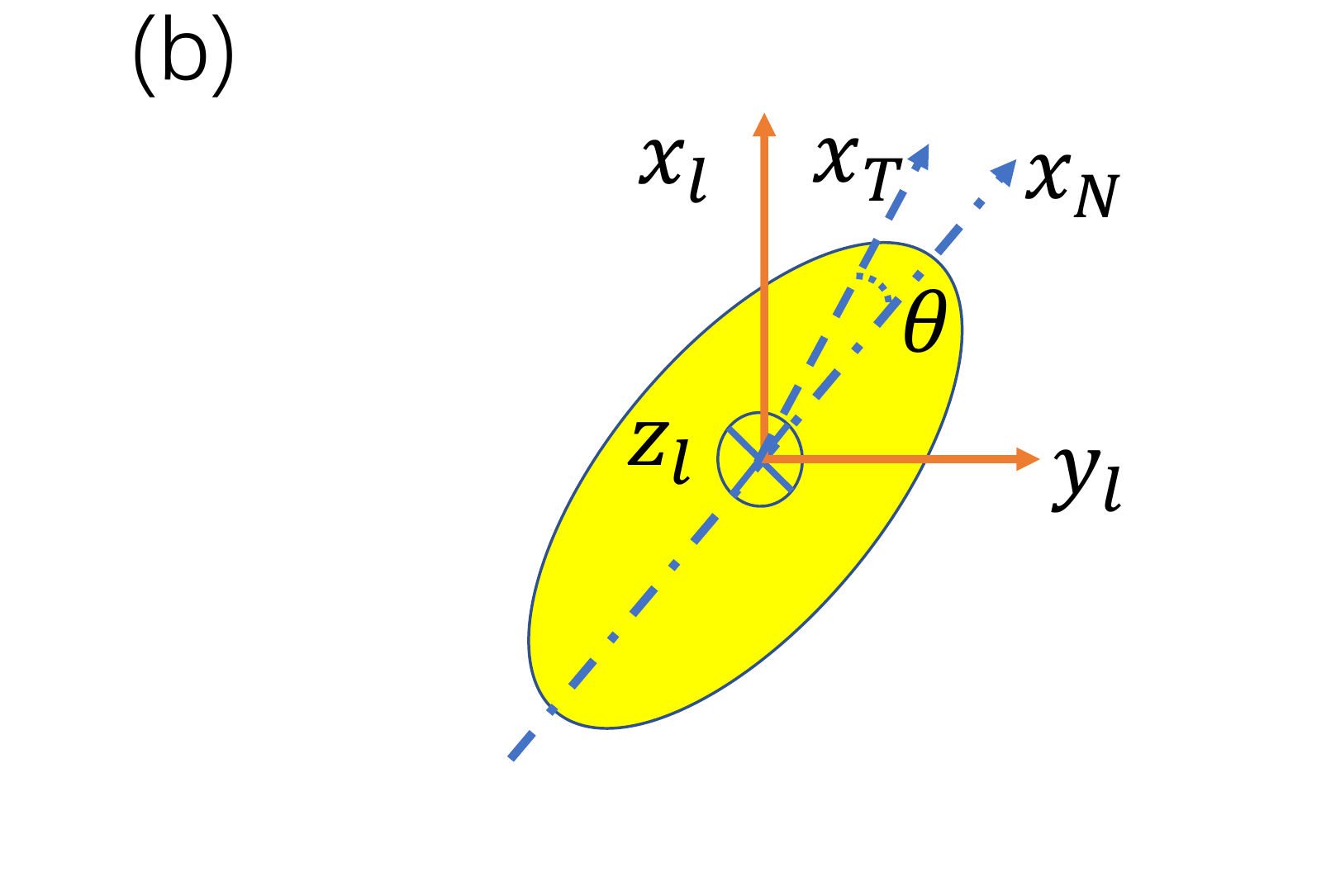}}
\caption{(a) A schematic diagram about a ellipsoidal nanoparticle trapped by a laser. (b) The relation between the direction of
systems of the nanoparticle ($x_N$), the trapping laser polarization ($x_T$), and the lens ($x_l$, $y_l$, $z_l$). The $x_N$ axis aligns with
the longest axis of the nanoparticle. $x_T$ and $x_l$ axes align with the trapping laser and the center of the two lenses,
respectively. The angle between $x_N$ and $x_T$ is $\theta$. The nanoparticle is trapped in a focal plane.}
\label{fig:experiment}
\end{figure}

As show in Fig.\ref{fig:experiment}, we consider one librational mode of an optically levitated ellipsoidal nanoparticle with long axis $r_{a}$, short axis $r_{b}=r_{c}$ and the density $\rho$. For simplicity, we suppose that all other degrees of freedom are frozen out.
The potential energy of the ellipsoid in the optical tweezers is~\cite{HoangPrl2016}:

\begin{equation}
U(\theta)=-\frac{V}{2c}[\kappa_{x}-(\kappa_{x}-\kappa_{y})\sin^2\theta]I_{0},
\label{Eq:the potential of tweezers}
\end{equation}
where $I_{0}$ is the intensity of trapping laser, $V=4\pi r_{a}r_{b}^{2}/3$ is the volume of the ellipsoid, $c$ is the speed of the light, $\kappa_{x,y}=\alpha_{x,y}/(\epsilon_{0}V)$ are the effective susceptibility of the ellipsoid, $\epsilon_{0}$ is the vacuum permittivity, and $\theta$ is the angle between the long axis($r_{a}$) of the ellipsoid and the electric field of the laser beam.

The Hamiltonian of the librational mode of this system is
\begin{equation}
H=T+U(\theta)
\label{Eq:initial hamiltonian},
\end{equation}
where $T=I\dot{\theta}^2/2$, with $I=4\pi\rho r_{a} r_{b}^2(r_{a}^2+r_{b}^2)/15$ being the rotational inertia of the ellipsoid.
By expanding Eq.(\ref{Eq:initial hamiltonian}) around the equilibrium position $\theta=0$ to the forth order, we get the effective Hamiltonian
\begin{equation}
H=\frac{I\dot{\theta}^2}{2}+\frac{I_0\text{V}(\kappa_{x}-\kappa_{y})}{2c}\theta^2-\frac{I_0\text{V}(\kappa_{x}-\kappa_{y})}{6c}\theta^4-\frac{I_0\text{V}\kappa_{x}}{2c}.
\label{Eq:Hamiltonian Taylor expansion}
\end{equation}
We have checked the sixth order term in the expansion of Eq. \eqref{Eq:initial hamiltonian}, and found that it is much smaller than the forth order term. Therefore,
 expanding Eq. \eqref{Eq:initial hamiltonian} to  the forth order is enough for our investigation. With the standard quantization procedure, the Hamiltonian (~\ref{Eq:Hamiltonian Taylor expansion}) turns into
\begin{eqnarray}
\hat{H}_{mec}&=&\hat{H_{0}}+\hat{H_{1}}=\hbar\omega_{t}^{\theta}\Big(\hat{b}^\dag\hat{b}+\frac{1}{2}\Big)-\hbar\eta\Big(\hat{b}^\dag+\hat{b}\Big)^{4}\nonumber\\
\text{with}~~ \omega_{t}^{\theta}&=&\sqrt{\frac{10P_{0}(\kappa_{x}-\kappa_{y})}{\pi w_{0}^{2}c\rho(r_{a}^{2}+r_{b}^{2})}},  ~~\eta=\frac{\hbar}{24I}
\label{Eq:Hamiltonian simplified}.
\end{eqnarray}
Here $\omega_{t}^{\theta}$ is the trapping frequency of the librational mode, $w_{0}$ is the waist of optical tweezers, the laser intensity is $I_{0}=2P_{0}/(\pi w_{0}^{2})$. The librational mode is described by the creation and annihilation operators, $\hat{b}^{\dag}$ and $\hat{b}$, which relate to the angle operator $\hat{\theta}$ and the angular momentum $\hat{J}_{\theta}$ by
 \begin{eqnarray}
\begin{aligned}
\hat{\theta}=\frac{\theta_{0}}{2}(\hat{b}+\hat{b}^{\dag})\\
\hat{J}_{\theta}=\frac{J_{0}}{2i}(\hat{b}-\hat{b}^{\dag})
\end{aligned}
\label{Eq:b phi },
\end{eqnarray}
where $\theta_{0}=\sqrt{\frac{2\hbar}{I\omega_{t}^{\theta}}}$ and $J_{0}=\sqrt{2I\hbar\omega_{t}^{\theta}}$.

\begin{figure}
\centering
\subfigure{
\begin{minipage}{0.45\textwidth}
\includegraphics[width=1\textwidth]{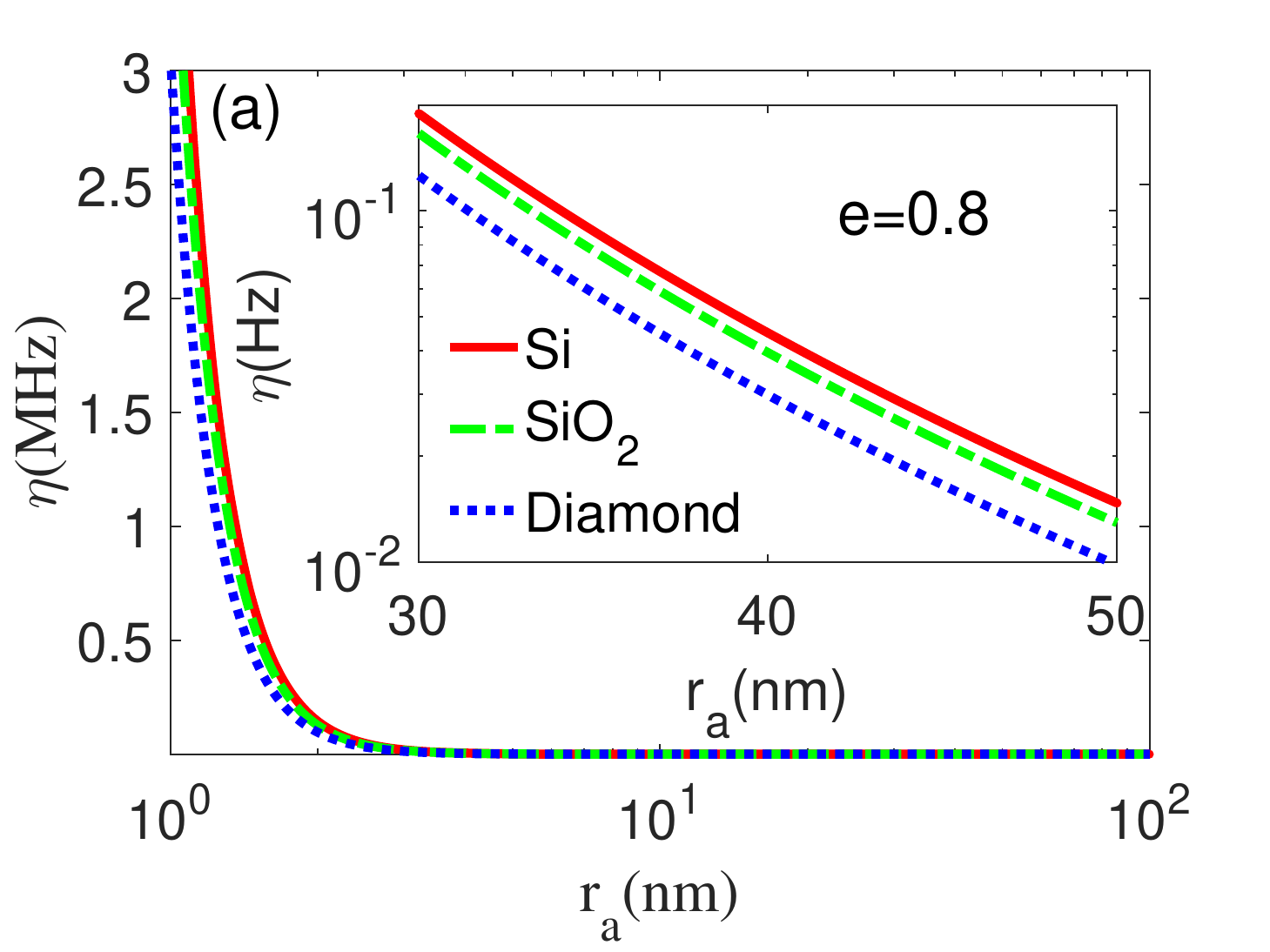}
\label{fig:I and eta B}
\end{minipage}}
\subfigure{
\begin{minipage}{0.45\textwidth}
\includegraphics[width=1\textwidth]{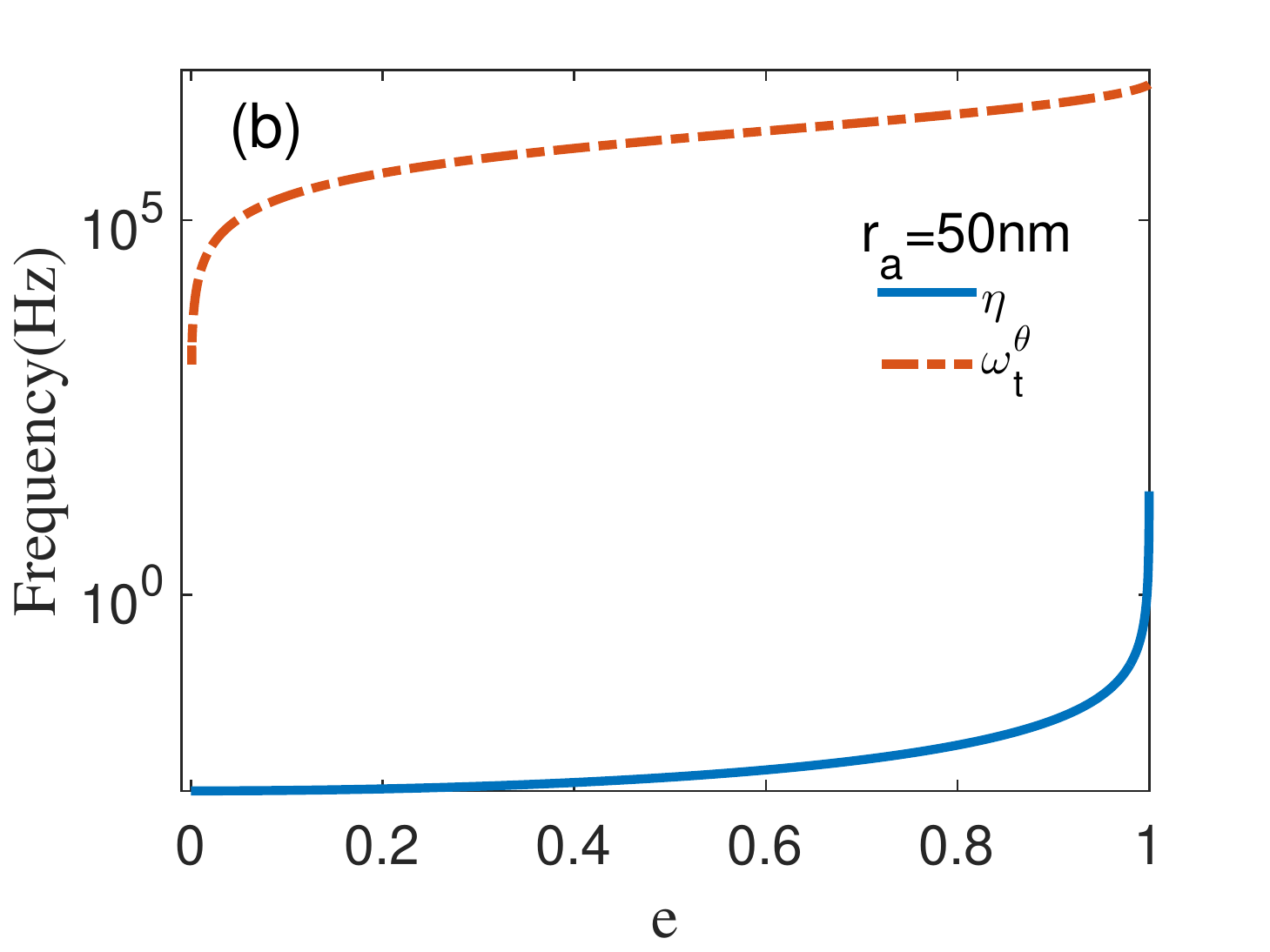}
\label{fig:I and eta C}
\end{minipage}}
\caption{The nonlinearity versus particle size, material, and eccentricity under the laser power $P_{0}=0.1~W$. (a) The nonlinearity versus long axis($r_{a}<100~nm$) of different material in $e=\sqrt{1-r_b^2/r_a^2}=0.8$. The inner figure shows the nonlinearity for different material.
(b) The nonlinearity and the trapping frequency of the diamond particle with different eccentricity which the long axis of nanoparticle is $r_{a}=50~nm$.}
\label{fig:I and eta}
\end{figure}

Equation.(~\ref{Eq:Hamiltonian simplified}) shows that the nonlinearity of the librational mode is inversely proportional to the inertia of ellipsoidal nanoparticle, but independent of the power and the shape of the optical trap.  The material, size and shape of nanopartilce determine the nonlinearity of the librational mode. As shown in Fig.~\ref{fig:I and eta B}, the nonlinearity is in inverse proportion to the fifth power of size, and in inverse proportion to the density of the particle. Furthermore, the nonlinearity increases with the increment of the eccentricity of the particle with the determined long axis. The shorter the long axis and the smaller the eccentricity can induce the larger nonlinearity.
For example, if the power of laser ($P_{0}=0.1~W$) and the particle's eccentricity($e=0.8$) is determined, the ratio between $\eta$ and $\omega_{t}^{\theta}$ are $4.6\times10^{-9}$ ($4.6\times10^{-5}$) for different long axis $50~nm$ ($5~nm$).
 The nonlinearity strength could be in the order of kHz when the long axis of the nanoparticle is around $5$ nm. If we reduce the power of the laser to decrease the trapping
 frequency in the order of $10$ kHz, the ratio ($\eta/\omega_{t}^{\theta}$) could approach $0.1$.  As the nonlinearity strength is comparable with the trapping frequency and  much larger than the librational mode decay, we may use the librational mode as a qubit. When the long axis is in the region $(10,~100)$~nm,  the nonlinearity of librational mode is at least two orders of magnitude larger than the nonlinearity of center-of-mass mode. A very interesting nonlinear phenomena like bistability may appear in this system.

\section{Bistability}
\label{SectIII}

Bistability is a ubiquitous phenomena in the nonlinear system. To exhibit the bistable effect, we drive the librational mode of this nonlinear system by a Gaussian laser beam, which linearly polarized at an angle $\pi/4$ with respect to the $x$ axis and modulated by $\sqrt{\cos(\omega_{ml}t)}$ in the direction of $x$ and $y$. It causes the vibrating amplitude and frequency of particle are $\Omega=\frac{P_{ml}V(\kappa_{x}-\kappa_{y})}{\pi\omega_{0}^{2}c}\sqrt{\frac{2}{\hbar I\omega_{t}^{\theta}}}$ and $\omega_{ml}$, where $P_{ml}$ is the power of the manipulative laser beam.  The Hamiltonian of librational mode under drive is
\begin{eqnarray}
 \begin{aligned}
 \hat{H}_{tot}&=\hat{H}_{mec}+\hat{H}_{dr},\\
 \hat{H}_{dr}&=\frac{\hbar\Omega}{2}(\hat{b}\mathrm{e}^{\mathrm{i}\omega_{ml}t}+\hat{b}^{\dag}\mathrm{e}^{-\mathrm{i}\omega_{ml}t}).
 \label{Eq:driving librational mode}
 \end{aligned}
\end{eqnarray}

In rotating wave frame, the Hamiltonian is transformed following $\hat{H}_{RM}=\hat{U}^{\dag}\hat{H}_{tot}\hat{U}-\hbar\omega_{ml}\hat{b}^{\dag}\hat{b}$ and $\hat{U}=\mathrm{e}^{-\mathrm{i}\omega_{ml}\hat{b}^{\dag}\hat{b}t}$,
\begin{eqnarray}
\hat{H}_{RM}=-\hbar\Delta_{ml}\hat{b}^{\dag}\hat{b}+\frac{\hbar\Omega}{2}(\hat{b}+\hat{b}^{\dag})-\hbar\eta(\hat{b}\mathrm{e}^{\mathrm{i}\omega_{ml}t}+\hat{b}^{\dag}\mathrm{e}^{-\mathrm{i}\omega_{ml}t})^{4}
\label{Eq:Hamiltonian in RWF},
\end{eqnarray}
where $\Delta_{ml}=\omega_{ml}-\omega_{t}^{\theta}$ is the detuning between the driving and the trapping laser beam.

The master equation can be used to describe the full dynamics of this nanoparticle which couples with the thermal bath~\cite{Louisell1973quantum},
\begin{eqnarray}
\dot{\hat{\rho}}(t)=\frac{1}{\mathrm{i}\hbar}[\hat{H}_{RM}(t),\hat{\rho}]+\mathscr{L}_{b}\hat{\rho}
\label{Eq:master equation},
\end{eqnarray}
where $\mathscr{L}_{b}=\frac{(1+\overline{n}_{b})}{2}\gamma_{b}\mathscr{D}_{b}+\frac{\overline{n}_{b}}{2}\gamma_{b}\mathscr{D}_{b^{\dag}}$, with the Lindblad operator $\mathscr{D}_{x}(\rho)=2x\rho x^{\dag}-x^{\dag}x\rho-\rho x^{\dag}x$. Here $\gamma_{b}$ is the decay of librational mode,  and $\overline{n}_{b}$ is the average phonon number of the thermal reservoir.

Using the master Eq.(~\ref{Eq:master equation}), we can deduce the motional equations of the librational mode mean field amplitudes $\beta$ ($\beta^{*}$)
\begin{equation}
\frac{\partial}{\partial t}
\begin{pmatrix}
\beta\\ \beta^{\ast}
\end{pmatrix}=
\begin{pmatrix}
\Big(\mathrm{i}\Delta_{ml}-\frac{\gamma_{b}}{2}+12\mathrm{i}\eta(|\beta|^{2}+1)\Big)\beta-\mathrm{i}\frac{\Omega}{2}\\
\Big(-\mathrm{i}\Delta_{ml}-\frac{\gamma_{b}}{2}-12\mathrm{i}\eta(|\beta|^{2}+1)\Big)\beta+\mathrm{i}\frac{\Omega}{2}
\end{pmatrix}
\label{Eq:matrix of beta}.
\end{equation}
Here we have neglected the highly oscillating terms with frequencies of $\pm 2\omega_{ml}$ and $\pm4\omega_{ml}$, and applied the semiclassical approximation (i,e. neglecting terms $\left\langle\hat{b}^{\dag}\hat{b}^{2}\right\rangle-\left\langle \hat{b}^{\dag}\right\rangle\left\langle\hat{b}^{2}\right\rangle$), $\left\langle \hat{b}^{\dag}\hat{b}^{2}\right\rangle=\beta^{\ast}\beta^{2}$. This approximation requires fluctuation terms $\left\langle(\delta\hat{b})^{2}\right\rangle$ and $\left\langle\delta\hat{b}^{\dag}\delta\hat{b}\right\rangle$ are much less than $|\beta|^2$.
In order to study the stability of the system, the amplitude of librational mode $\beta(t)$ is split into two terms: an average amplitude $\beta_{0}$ and a fluctuation $\beta_{1}(t)$,
 \begin{equation}
 \beta(t)=\beta_{0}+\beta_{1}(t)
 \label{Eq:{steady and fluctuation}},
 \end{equation}
where $\beta_{0}$ is the steady state solution of Eq.(\ref{Eq:matrix of beta}), which satisfies
 \begin{equation}
   \begin{pmatrix}
    \frac{\mathrm{i}\Omega}{2}\\\frac{-\mathrm{i}\Omega}{2}
     \end{pmatrix}
     =
    \begin{pmatrix}
\Big(\mathrm{i}\Delta_{ml}-\frac{\gamma_{b}}{2}+12\mathrm{i}\eta(n+1)\Big)\beta_{0}\\
\Big(-\mathrm{i}\Delta_{ml}-\frac{\gamma_{b}}{2}-12\mathrm{i}\eta(n+1)\Big)\beta_{0}^{\ast}
   \end{pmatrix}.
   \label{Eq:steady state equation}
 \end{equation}
Here $n=|\beta_{0}|^{2}$ and the fluctuation $\beta_{1}$ satisfies
\begin{equation}
\frac{\partial}{\partial t}
\begin{pmatrix}
\beta_{1}(t)\\ \beta_{1}(t)^{\ast}
\end{pmatrix}=-\textbf{A}
\begin{pmatrix}
\beta_{1}(t)\\\ \beta_{1}^{\ast}(t)
\end{pmatrix}
 \label{Eq:fluctuation equation},
\end{equation}
and
\begin{equation}
\textbf{A}=
\begin{pmatrix}
\kappa+2\chi n && \chi\beta_{0}^{2}\\
\chi^{\ast}\beta_{0}^{\ast 2} && \kappa^{\ast}+2\chi^{\ast}n
\end{pmatrix}
\label{Eq:A}.
\end{equation}
For simplicity, we have defined that $\chi=-12\mathrm{i}\eta$ and $\kappa=\gamma_{b}/2-\mathrm{i}(\Delta_{ml}+12\eta)$.

\begin{figure}[htbp]
  \centering
  \includegraphics[width=0.48\textwidth]{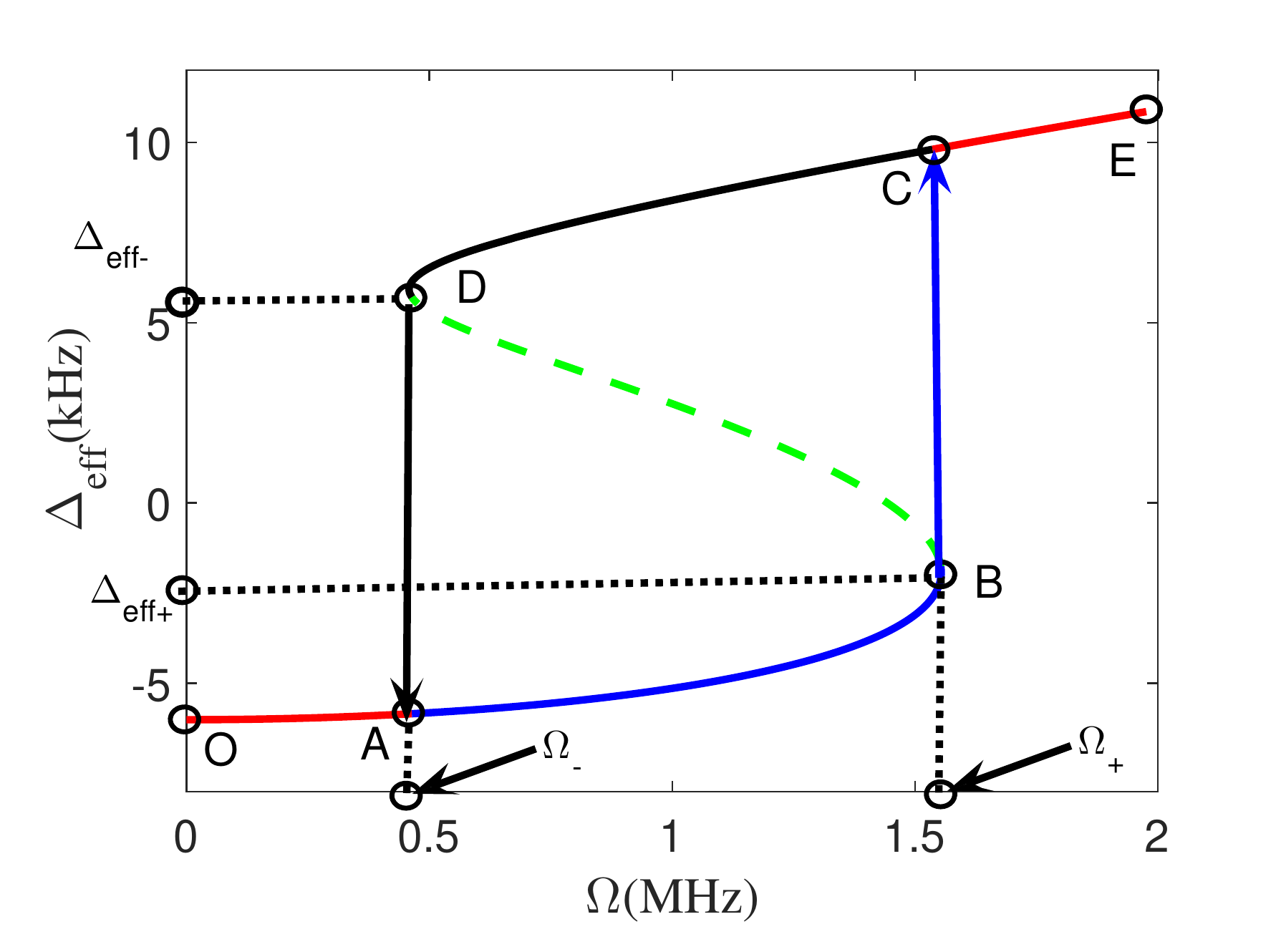}
  \caption{Bistability of this system in $P=10~m\text{Torr}$.  The figure is the effect detuning $\Delta_{eff}$ versus driven amplitude $\Omega$. There are several characteristic points, O(0, -6.00~kHz), A(466~kHz, -5.8~kHz), B(1.55~MHz, -1.65~kHz), C(1.55~MHz, 9.86~kHz), D(466~kHz,5.89~kHz), E(2.00~MHz, 10.9~kHz), and the detuning $\omega_{ml}-\omega_{t}^{\theta}=-6007~\text{Hz}$. The turning points are $B(\Omega_{+}, \Delta_{eff+})$ and $C(\Omega_{-}, \Delta_{eff-})$ respectively. In this case, the particle with the long axis $r_{a}=50nm$ and $e=0.9$ is trapped by $P=0.1W$ and $w_{0}=0.6\mu m$ laser. The temperature of the system is $T=300~K$ in this all paper. }
  \label{fig:bistability sigma401}
\end{figure}

In order to obtain stable eigenvalues we calculate
$\text{Tr}(\textbf{A})=\gamma_{b}>0$ and $\text{Det}(\textbf{A})=|\kappa|^2+2(\chi\kappa^{\ast}+\kappa\chi^{\ast})n+3|\chi|^{2}n^{2}$.
Using the Routh-Hurwitz criterion \cite{DeJesusPra1987}, it's found that Eq.(\ref{Eq:fluctuation equation}) has steady solution when $\text{Tr}(\textbf{A})>0$ and $\text{Det}(A)>0$.
For the specific $n$, the Routh-Hurwitz criterion gives stable eigenvalues. The stable condition implies
\begin{equation}
0\leq\omega_{ml}<\omega_{t}^{\theta}-12\eta-\frac{\sqrt{3}\gamma_{b}}{2}
\label{Eq:the region of bistability}.
\end{equation}
From Eq.(\ref{Eq:the region of bistability}), we know that the bistability only appears in the red detuning ($\Delta_{ml}<0$) regime, and the blue-detuning drive will not cause bistability.
For convenience, we set $\Delta_{eff}=\Delta_{ml}+24\eta n$, $\omega_{c}=\omega_{t}^{\theta}-\delta_{c}$, and  $\delta_{c}=12\eta(\zeta+1)$, where $\zeta={\sqrt{3}\gamma_{b}}/{24\eta}$. If $\omega_{ml}=\omega_{t}^{\theta}-12\eta(\zeta+1)+\delta$, $\delta$ varies about $\omega_{c}$, From Eq.(\ref{Eq:steady state equation})and (~\ref{Eq:the region of bistability}) we can get
\begin{widetext}
\begin{eqnarray}
\begin{aligned}
\Omega^{2}&=\Big(\gamma_{b}^{2}+\big(\Delta_{eff}+\delta-12\eta(\zeta-1)\big)^{2}\Big)\frac{\Delta_{eff}-\delta+12\eta(\zeta+1)}{24\eta} \label{Eq:steady equation Delta eff2},\\
\Delta_{eff\pm}&=\frac{-\delta+12\eta(\zeta-3)\pm2\sqrt{\delta^{2}-\sqrt{3}\gamma_{b}\delta}}{3}
\label{Eq:turning point Delta eff2}.
\end{aligned}
\end{eqnarray}
\end{widetext}
From Eq.(~\ref{Eq:turning point Delta eff2}), we find that when driven frequency($\omega_{ml}$) is very low, the system needs the strong driven amplitude($\Omega$) for bistability. However, if the driving frequency is near the resonant frequency, a very weak driving amplitude can also induce bistability. Obviously, the distance between two turning points $\Delta_{eff\pm}$ is
\begin{equation}
\Delta_{w}=\Delta_{eff+}-\Delta_{eff-}=\frac{4\sqrt{\delta^{2}-\sqrt{3}\delta\gamma_{b}}}{3}
\label{Eq:dispalcement of turning points},
\end{equation}
which only depends on detuning and environment. When $\delta=\delta_{0}=12\eta(\zeta-1)-\Delta_{eff}$, $\Omega$ have the minimum $\gamma_{b}\sqrt{\frac{\Delta_{eff}+12\eta}{12\eta}}$.
 This line is constituted by one turning point of the bistable state, as shown like the point D in Fig.\ref{fig:bistability sigma401}. It's easy to know that the output ($\Delta_{eff}$) is parabola form about driven amplitude ($\Omega$). As a matter of fact, every point in this line is one of the jump points in different bistable state.

 The bistability of the librational mode is shown in Fig.~\ref{fig:bistability sigma401}. The lines $OABCE$ and $ECDAO$ show two stable paths. The dashed line $DB$ is unstable and will not appear in experiments. When $\Delta_{eff}$ increases with the increment of the amplitude ($\Omega$) from $O$, $\Delta_{eff}$ will jump from $B$ to $C$ in the turning point of $B$, and then increases to $E$. On the other hand, if $\Omega$ decreases from $E$, $\Delta_{eff}$ should jump down from the turning point $D$ to $A$ with the decrement of $\Omega$ and then decrease to $O$. In brief, the driving frequency ($\omega_{ml}$) determines the system whether happens bistability, and the driving amplitude determines that which state the system will be go though.

\begin{figure}[htbp]
  \centering
  \includegraphics[width=0.48\textwidth]{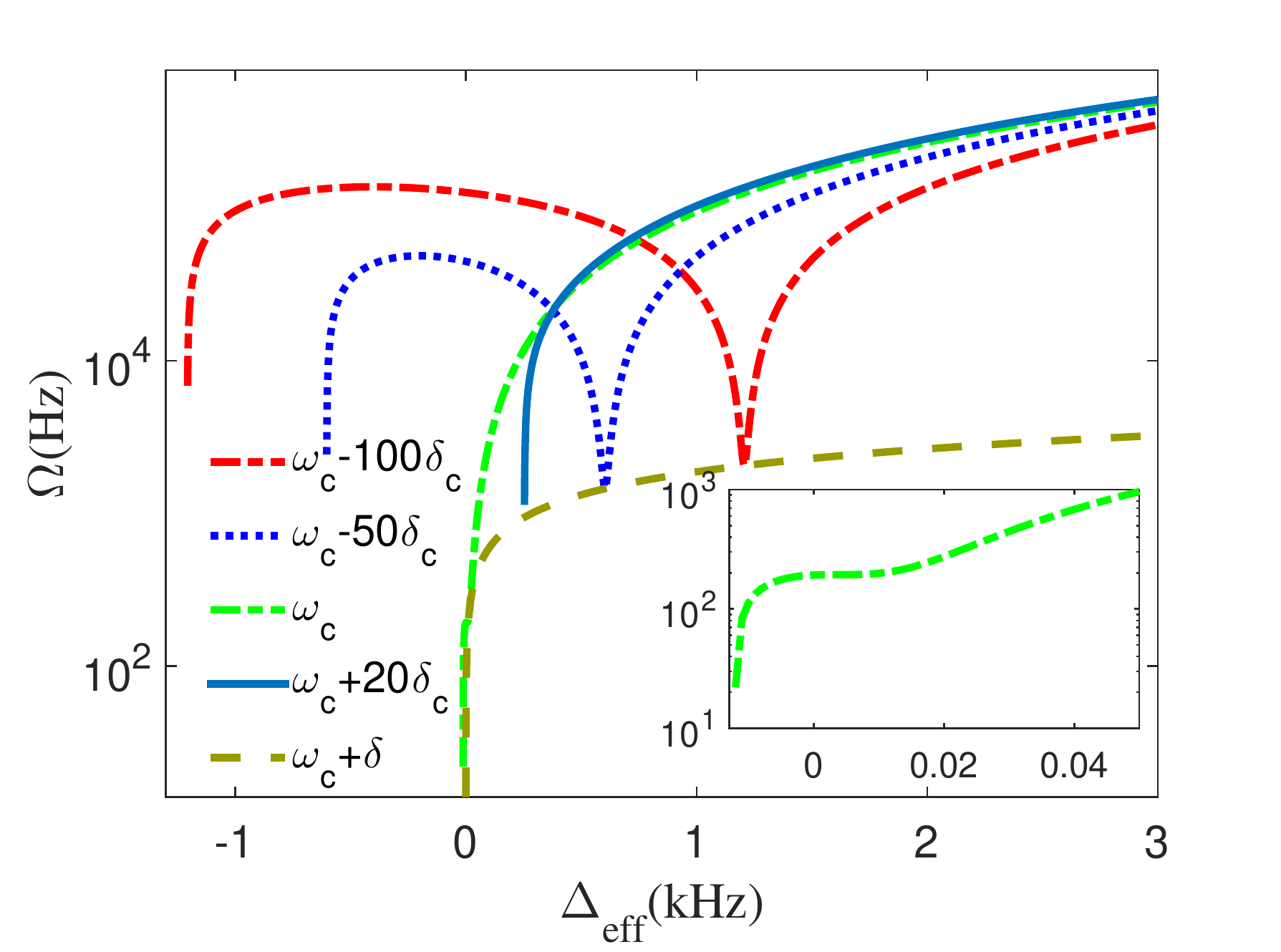}
  \caption{Bistability of this system for different driving frequency($\omega_{ml}$).  The figure is the driven amplitude $\Omega$ versus effect detuning $\Delta_{eff}$, different color line presents different driven frequency in determined resident air pressure $P_{air}=1~\mu\text{Torr}$. Every line has two extreme value which are the turning points and the distance of them is the bistable window. The small figure is steady-state for the critical driving frequency($\omega_{c}$).}
  \label{fig:bistability Delta}
\end{figure}

Effective detuning is an important physical quantity. Fig.~\ref{fig:bistability Delta} shows the the response of effect detuning ($\Delta_{eff}$) about the driven amplitude ($\Omega$). When $\omega_{ml}<\omega_{c}$, bistability will appear, and when $\omega_{ml}>\omega_{c}$, system has no bistability.
When $\omega_{ml}=\omega_{c}$, a platform will appear as shown in the inner figure of Fig.~\ref{fig:bistability Delta}. The width of the window is determined by detuning of driving frequency. The larger the driving detuning $\Delta_{ml}$, the more the distance $\Delta_w$ between two turning points.

\begin{figure}
\centering
\subfigure{
\begin{minipage}{0.45\textwidth}
\includegraphics[width=1\textwidth]{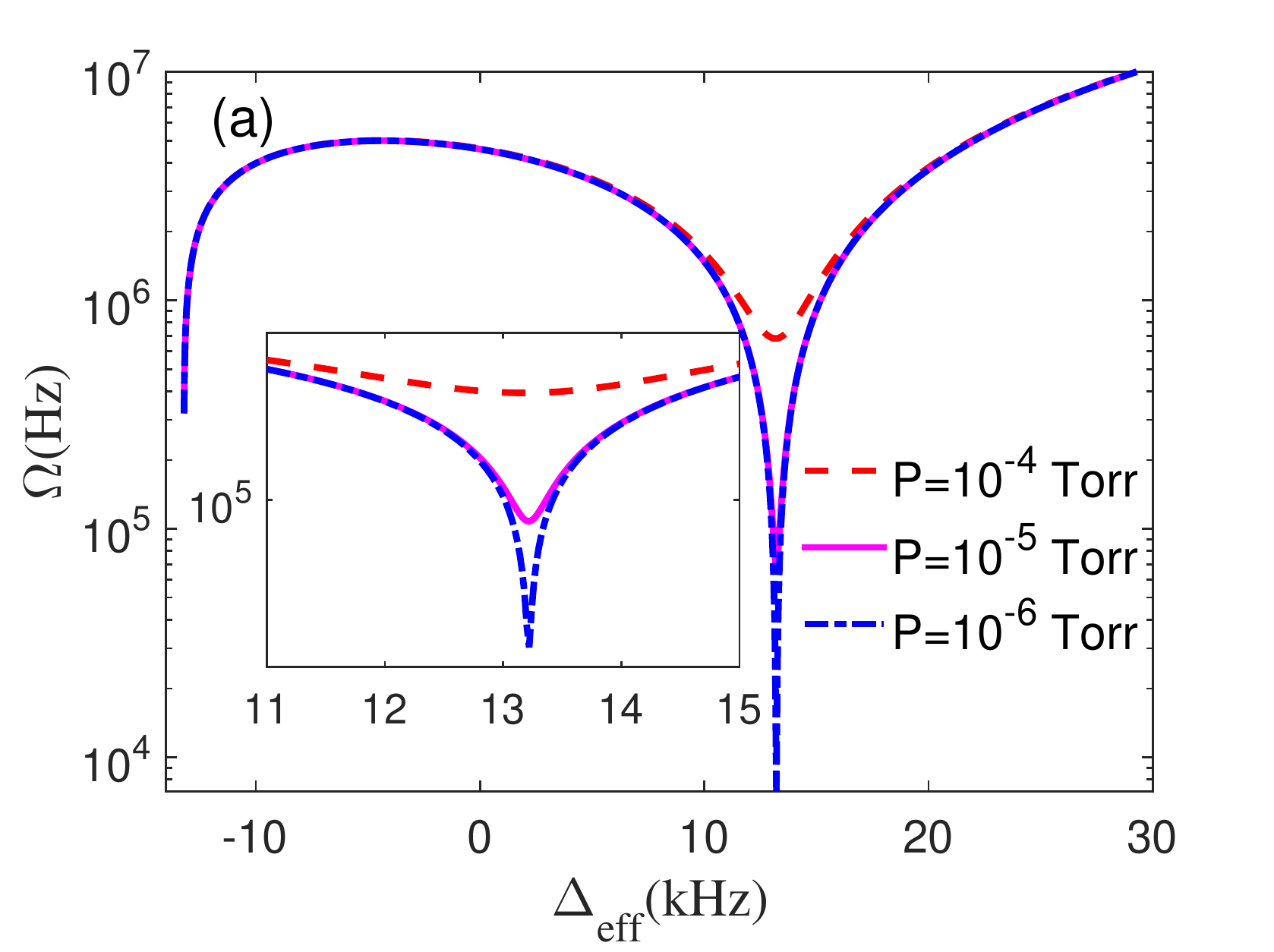}
\label{fig:bistability different pressure}
\end{minipage}}
\subfigure{
\begin{minipage}{0.45\textwidth}
\includegraphics[width=1\textwidth]{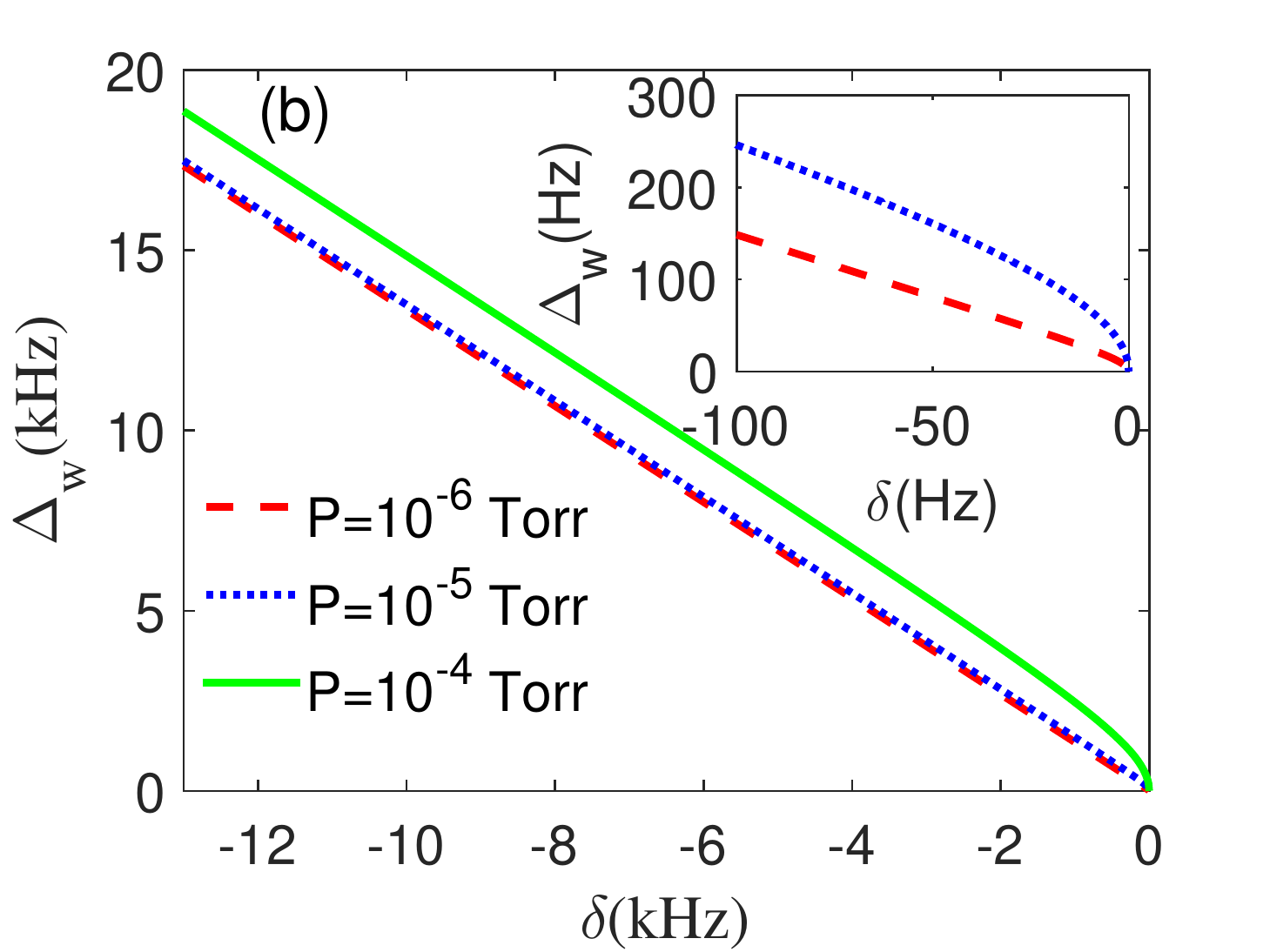}
\label{fig:displacement between turning points}
\end{minipage}}
 \caption{Bistability of this system and bistable windows in different residual air pressure. (a) The figure is the driven amplitude $\Omega$ versus detuning $\Delta_{eff}$, different line present different resident air pressure in determined driven frequency $\omega_{ml}-\omega_{t}^{\theta}=-13.26~\text{kHz}$. (b) The figure is the breadth of bistability $\Delta_{w}$ versus $\delta$, different line present different resident air pressure. The small figure is the case of $P=10^{-5}~Torr$ and $P=10^{-6}~Torr$.}
\label{fig:bistability air pressure}
\end{figure}

If the driven frequency is determined, the relation between $\Omega$ and $\Delta_{eff}$ will be affected only by the residual air pressure ($P$) and temperature ($T$). For the fixed experimental temperature, the residual air pressure only affect the width of bistable window. In Fig.~~\ref{fig:bistability different pressure}, the width of window changes little for different $P$, but the depth is inverse proportional with the air pressure. That is because the proportional relation between $P$ and $\gamma_{b}$. The larger the decay $\gamma_{b}$, the more energy will be need for bistability, so the driven amplitude will be bigger. And  Fig.~\ref{fig:displacement between turning points} shows the relation between the bistable window $\Delta_{w}$ and characteristic detuning $\delta$ in different pressure. The higher the air pressure, the larger the  window width. When the residual air pressure is very low
($P<10^{-5}\text{Torr}$), the $\Delta_{w}$ will not change any more. It's worth to mention that $\Delta_{w}$ is closed when characteristic detuning is zero.

\section{squeezing}
\label{SectIV}
In the last section, we discussed the bistability of the librational mode. Here, we discuss another nonlinearity induced phenomena, librational mode squeezing.
In order to generate squeezed state of the librational mode, we choose the driving detuning and strength to fulfill the stable conditions.
By standard linearisation method, we can set $\hat{b}\rightarrow\beta_{0}+\delta\hat{b}$, $\beta_{0}$ is steady state amplitude of librational mode, and $\delta\hat{b}$ is fluctuation around steady state, for simplicity, $\delta\hat{b}$ is written as $\hat{b}$.
Therefore, under the rotating wave approximation and setting $\beta_{0}=r\mathrm{e}^{\mathrm{i}\phi}$, $r\in\mathbb{R}$  also $|\beta_{0}|\gg 1$, the linearized Hamiltonian reads
\begin{equation} \label{Eq:linearisated Hamiltonian}
\hat{H}_{l}=-\hbar(\Delta_{ml}+24\eta r^{2})\hat{b}^{\dag}\hat{b}-6\hbar\eta r^{2}(\mathrm{e}^{2\mathrm{i}\phi}\hat{b}^{\dag 2}+\mathrm{e}^{-2\mathrm{i}\phi}\hat{b}^{2}).
\end{equation}
This Hamiltonian can apply for single-mode squeezing.

Base on Eq.(~\ref{Eq:b phi }), we introduce the Hermitian amplitude operators $\hat{X}_{1}$ and $\hat{X}_{2}$ which are essentially dimensionless angle $\hat{\theta}$ and angular momentum $\hat{J}_{\theta}$ operators,
$\hat{X}_{1}=\hat{\theta}/\theta_{0}$ and
$\hat{X}_{2}=\hat{J}_{\theta}/J_{0}$. We define the variances, $\mathcal{S}_{\theta}=\Delta X_{1}=[\langle\hat{\theta}^{2}\rangle-\langle\hat{\theta}\rangle^{2}]/\theta_{0}^{2}$ and $\mathcal{S}_{J}=\Delta X_{2}=[\langle\hat{J}_{\theta}^{2}\rangle-\langle\hat{J}_{\theta}\rangle^{2}]/J_{0}^{2}$. When $\mathcal{S}_{\theta(J)}$ satisfies the relation
\begin{equation}
\mathcal{S}_{\theta(J)}^{2}<\frac{1}{4}
\label{Eq:D Xi}
\end{equation}
in the vacuum state, we call $\theta(J_{\theta})$ is squeezed, which is useful for the precision measurement of angle and angular momentum. We will discuss how to squeeze the thermal and vacuum librational mode.

In experiments, usually, the initial state is not vacuum, but is in the thermal equilibrium. The state of a system with Hamiltonian $\hat{H}$ is represented by the density matrix
\begin{equation}
\hat{\rho}_{th}=\frac{\mathrm{exp}(-\beta_{th}\hat{H})}{\mathrm{Tr}[\mathrm{exp}(-\beta_{th}\hat{H})]}
\label{Eq:thermal state}
\end{equation}
where $\beta_{th}=(k_{B}T)^{-1}$, $k_{B}$ is Boltzmann's constant and $T$ is temperature. It's well known that the average of the number operator in thermal state is mean thermal phonon number $\bar{n}$. We can use the squeezing Hamiltonian Eq. \eqref{Eq:linearisated Hamiltonian} to generate the thermal squeezing state \cite{Rashid2016}.
The variance of $\hat{X}_{1}$ and $\hat{X}_{2}$ in this state will be got
as follow
\begin{widetext}
\begin{eqnarray}
\mathcal{S}_{\theta}(t)=\frac{\langle2\hat{b}^{\dag}(0)\hat{b}(0)+1\rangle}{4\lambda_{p}^{2}}&\times&\Big((\xi^{2}-\lambda\xi\cos(2\phi))\sinh(2\lambda_{p}t)-\lambda\xi\sin(2\phi)\cosh(2\lambda_{p}t)\nonumber\\
&-&(\lambda^{2}-\lambda\xi\cos(2\phi))\Big)
\label{Eq:Vx1},
\end{eqnarray}
\end{widetext}
where $\lambda=\Delta_{ml}+24\eta r^{2}$, $\xi=12\eta r^{2}$ and $\lambda_{p}=\sqrt{\xi^{2}-\lambda^{2}}$. Meanwhile, the variance of $X_{2}$ evolves as:
\begin{widetext}
\begin{eqnarray}
\mathcal{S}_{J}(t)=\Big(\big(\xi^4\sin^2(2\phi)&+&\lambda_{p}^2\xi^2-\lambda\xi^3\sin^{2}(2\phi)\cos(2\phi)+\lambda_{p}^2\lambda\xi\cos(2\phi)-2\lambda_{p}\lambda\xi^{2}\sin^{2}(2\phi)\big)\cosh(2\lambda_{p}t)\nonumber\\
&-&\big(\lambda\xi^2\sin^2(2\phi)+\lambda_{p}^2\lambda-\lambda_{p}\xi^{2}-\lambda_{p}\lambda\xi\cos(2\phi)\big)\sin(2\phi)\sinh(2\lambda_{p}t)\nonumber\\
&-&\lambda(\xi^{2}\sin^2(2\phi)-\lambda_{p}^{2})(\lambda-\xi\cos(2\phi))\Big)\times\frac{\langle2\hat{b}^{\dag}(0)\hat{b}(0)+1\rangle}{4\lambda_{p}^2(\xi\cos(2\phi)-\lambda)^2}
\label{Eq:Vx2}.
\end{eqnarray}
\end{widetext}

Here, the average of $\langle\hat{b}^{\dag}(0)\hat{b}(0)\rangle$ is the mean thermal phonon number $\bar{n}$. We can determine whether that $\Delta X_{\mathit{i}}(\mathit{i}=1,2)$ is squeezed in thermal state by \cite{Rashid2016}
\begin{equation}
(\Delta X_{\textit{i}})^{2}<\frac{1}{4}(2\bar{n}+1)
\label{Eq:thermal D Xi}.
\end{equation}
By using the squeezed thermal state, we may increase the measurement precision on either angle or angular momentum without cooling the system.
 If system is in the vacuum state $(\langle\hat{b}^{\dag}(0)\hat{b}(0)\rangle=0)$ at the initial time, this situation will reduce to Eq.(\ref{Eq:D Xi}).

It's easy to know that the formation of $\lambda_{p}$ decides the property of $\mathcal{S}_{\theta}$, now we must analysis the
property of $\lambda_{p}$. The system two characteristic points, $\omega_{ml1}=\omega_{t}^{\theta}-36\eta r^{2}$ and $\omega_{ml2}=\omega_{t}^{\theta}-12\eta r^{2}$.
At first, we consider $\omega_{ml1}<\omega_{ml}<\omega_{ml2}$, where $\lambda_{p}\in\mathbb{R}$ and $\lambda_{p}>0$. In this case, we can define $\cos(\arctan\frac{\lambda_{p}}{\lambda})=\frac{\lambda}{\sqrt{\lambda^{2}+\lambda_{p}^{2}}}$ and $\sin(\arctan\frac{\lambda_{p}}{\lambda})=\frac{\lambda_{p}}{\sqrt{\lambda^{2}+\lambda_{p}^{2}}}$, Eq.(\ref{Eq:Vx1}) can simplify as follow:
\begin{eqnarray}
\mathcal{S}_{\theta}=\frac{\xi^{2}}{4(\xi^{2}-\lambda^{2})}
&\times&\Big( \big(1\!-\!\cos(2\phi-\arctan\frac{\lambda_{p}}{\lambda})\big)\mathrm{e}^{2\lambda_{p}t}\nonumber\\
&+&\big(1\!-\!\cos(2\phi+\arctan\frac{\lambda_{p}}{\lambda})\big)\mathrm{e}^{-2\lambda_{p}t}\nonumber\\
&+&2\frac{\lambda}{\xi}\big(\cos(2\phi)-\frac{\lambda}{\xi}\big)\Big)
\label{Eq:simplify squeezing rate 1}.
\end{eqnarray}
 For different squeezing angle, we have different squeezing of variance $\mathcal{S}_{\theta}$ as follow:
 \begin{eqnarray}
 \mathcal{S}_{\theta}=
\begin{cases}
\frac{1}{4}\mathrm{e}^{-2\lambda_{p}t} & \phi=\frac{1}{2}\arctan(\frac{\lambda_{p}}{\lambda})+n\pi\\
\frac{1}{4}\mathrm{e}^{2\lambda_{p}t}  & \phi=n\pi-\frac{1}{2}\arctan(\frac{\lambda_{p}}{\lambda})\\
\end{cases}
\label{Eq:decay and rising squeezing rate }.
\end{eqnarray}
It is easy to know that when $\phi=\frac{1}{2}\arctan(\frac{\lambda_{p}}{\lambda})+n\pi$, $\mathcal{S}_{\theta}$ will decrease to zero in a particular direction, which present that the fluctuation of $\theta$ will be zero and the measurement accuracy is very high, and $\lambda_{p}$ determines the speed of decay. Namely, $\Delta_{ml}$ and $\eta r^{2}$ codetermine the decay speed. At the same time, another direction will increase undoubtedly.

The second case is $\omega_{ml}<\omega_{ml1}$ or $~\omega_{ml}>\omega_{ml2}$. Obviously, $\lambda_{p}^{2}<0$, so
$\lambda_{p}=\pm\mathrm{i}\lambda_{p}'$ and $\lambda_{p}'=\sqrt{(\omega_{ml}-\omega_{ml1})(\omega_{ml}-\omega_{ml2})}>0$, the squeezing of the variance of $\hat{X}_{1}$ will be
\begin{widetext}
\begin{eqnarray}
\mathcal{S}_{\theta}=\frac{1}{4\lambda_{p}^{2}}&\times&\Big( \xi\big(\xi-\lambda\cos(2\phi)\big)\cos(2\lambda_{p}'t)+\xi\lambda_{p}'\sin(2\phi)\sin(2\lambda_{p}'t)+\lambda\big(\xi\cos(2\phi)-\lambda)\big)\Big)
\label{Eq:simplify squeezing rate 2}.
\end{eqnarray}
\end{widetext}
 It is the same with the last subsection, different squeezing direction will give different variance squeezing. In some direction, the librational mode is amplified, for example $\phi=\frac{\pi}{2}+m\pi$, however, the librational mode is oscillatorily squeezed in $\phi=\pi+m\pi$,  and the others are the result of joint action of the amplification and squeezing. The oscillatory periods of them are the same, $\pi/\lambda_{p}'$.
 \begin{eqnarray}
 \mathcal{S}_{\theta}=
\begin{cases}
\frac{1}{4}-\frac{3\eta r^2}{\Delta_{ml}+24\eta r^{2}}\sin(2\lambda_{p}'t)& \phi=\frac{1}{2}\arccos(\frac{\xi}{\lambda})\\
\frac{1}{4}+\frac{3\eta r^2}{\Delta_{ml}+12\eta r^{2}}(1-\cos(2\lambda_{p}'t))  & \phi=\frac{\pi}{2}+m\pi\\
\frac{1}{4}+\frac{3\eta r^2}{\Delta_{ml}+36\eta r^{2}}(\cos(2\lambda_{p}'t)-1)  & \phi=\pi+m\pi\\
\frac{1}{4}-\frac{\xi\sin(\lambda_{p}^{'}t)}{2(\xi^2-\lambda^2)}(\xi\sin(\lambda_{p}'t)\mp\lambda_{p}^{'}\cos(\lambda_{p}'t))  & \phi=\pm\frac{\pi}{4}+m\pi\\
\end{cases}
 \label{Eq:oscillation squeezing rate}
 \end{eqnarray}

 \begin{figure}
\centering
\subfigure{
\begin{minipage}{0.48\textwidth}
\includegraphics[width=1\textwidth]{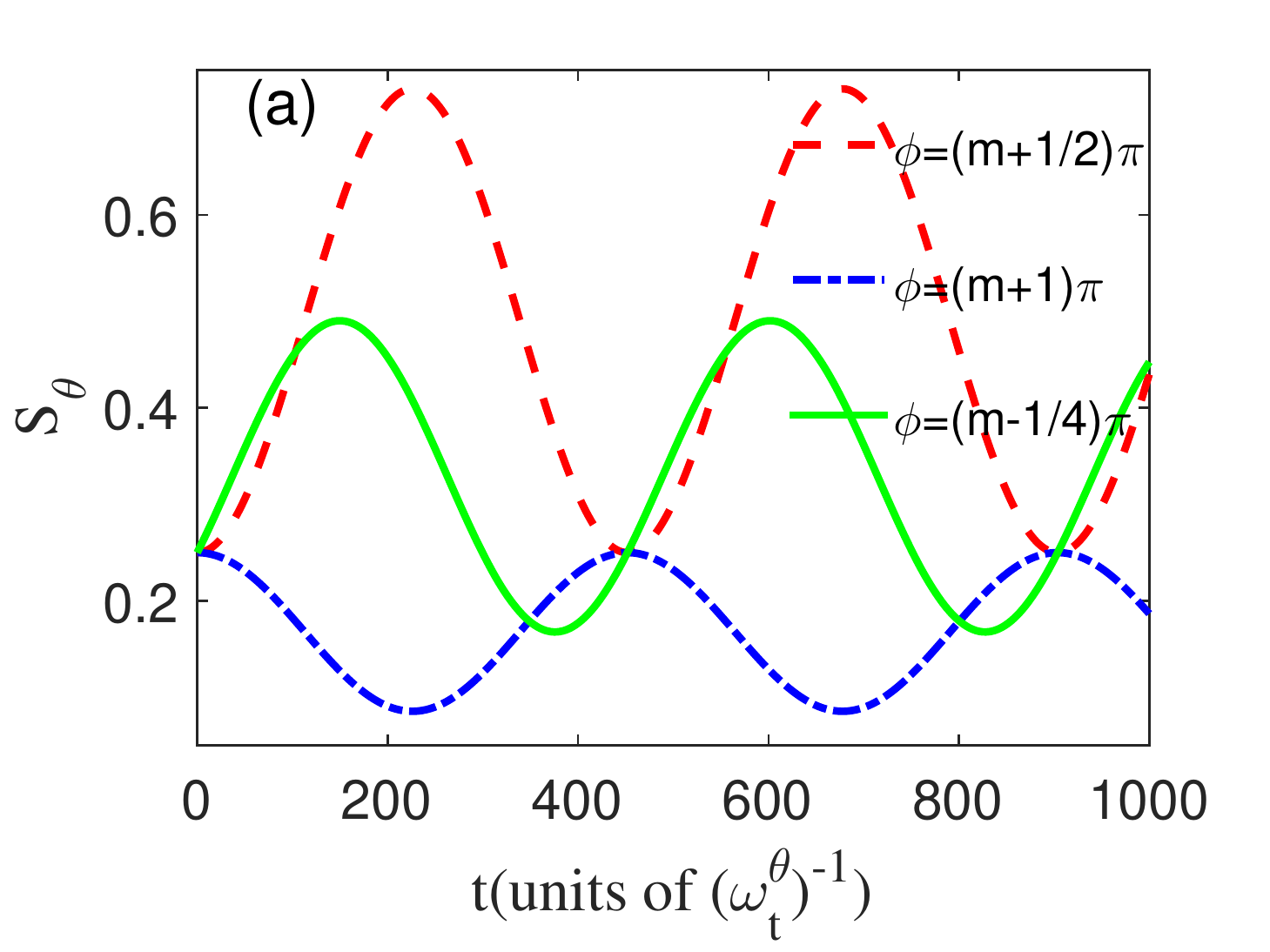}
\label{fig:oscillation squeezing}
\end{minipage}}
\subfigure{
\begin{minipage}{0.48\textwidth}
\includegraphics[width=1\textwidth]{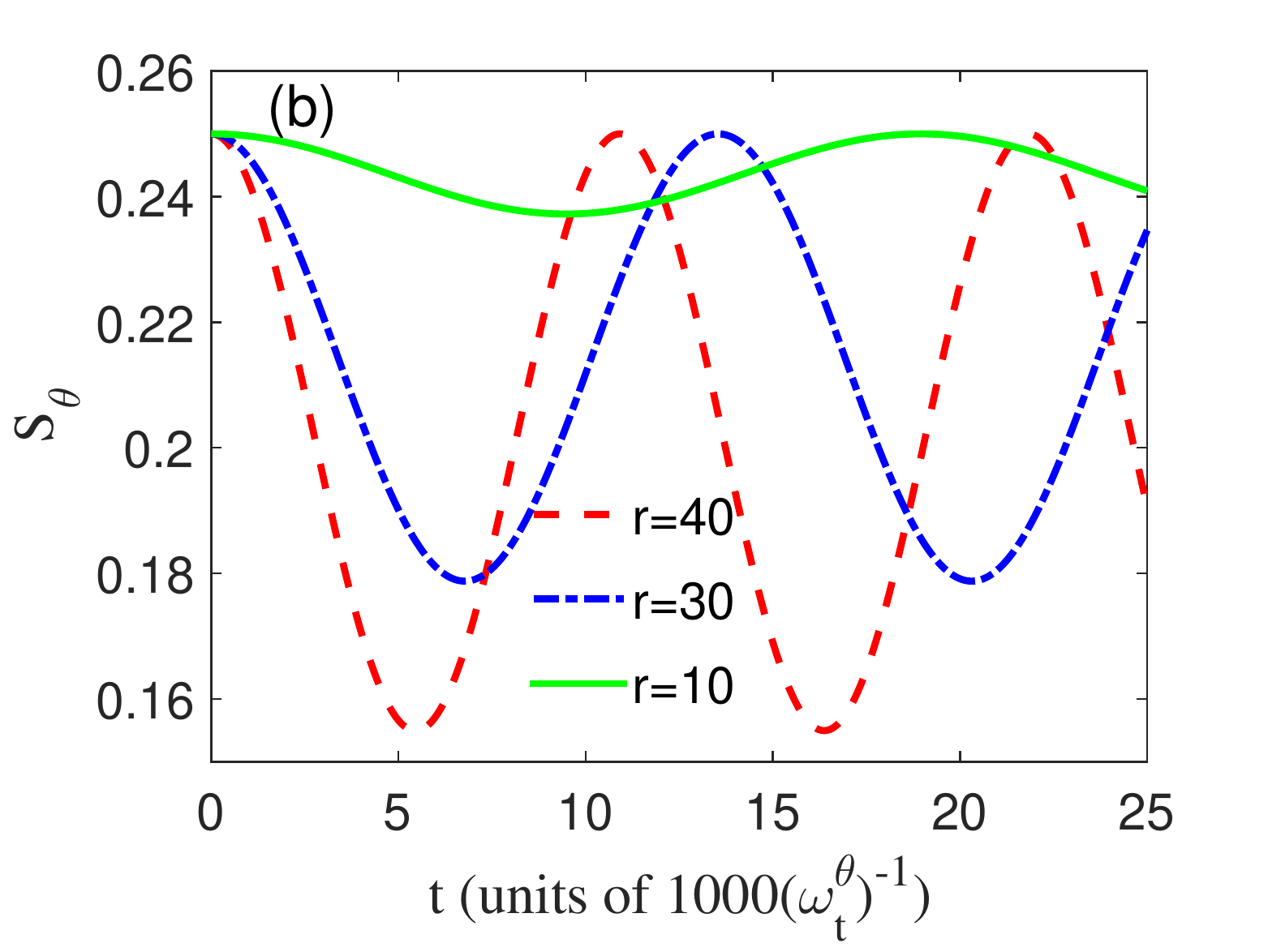}
 \label{fig:osscilation squeezing different R}
\end{minipage}}
 \caption{The oscillatory squeezing of the librational mode. The abscissas are respectively in the unit of the oscillatory period and the 1000 times oscillatory period of the librational mode. The oscillatory period of the librational mode is the $0.79\mu s$. (a) represents different variance squeezing of different direction in oscillation and $r=40$. These lines denote different squeezing direction. (b)  shows the oscillation squeezing $X_{1}$ of the direction of $\phi=(n+1)\pi$ in different $r$. Different line represents the behaviour different oscillatory squeezing in different squeezing amplitude. In this case, $r_{a}=50~nm$, $r_{b}=40~nm$, $\omega_{t}^{\theta}=1.2621\text{MHz}$, and $\omega_{ml}-\omega_{t}^{\theta}=0.2~\text{kHz}$. }
 \label{fig:oscillatory squeezing and amplifying}
\end{figure}

 In order to explain more clearly, we plot a Fig.~\ref{fig:oscillatory squeezing and amplifying}. From the Fig.~\ref{fig:oscillation squeezing}, when the squeezing phase angle is $\phi=(m+1)\pi$, the squeezing ratio of $\hat{X}_1$ is always less than $0.25$, namely the metrical uncertainty of the librational angle is decreased. However, other direction does not fully present squeezing property.
Therefore, in some direction we can measure always more accurately about $X_{1}$ or $\theta$ like $\phi=(m+1)\pi$ in Fig.~\ref{fig:oscillation squeezing} and Fig.~\ref{fig:osscilation squeezing different R}, on the contrary, the measurement of angular momentum is not so accurate.

At the same time, different squeezing amplitude can induce different squeezing ratio and squeezing oscillating period in the certain squeezing direction. For example, in the direction of $\phi=(m+1)\pi$, the squeezing ratio and the squeezing oscillating period is determined by the driving frequency($\omega_{ml}$) and the squeezing amplitude($r$). The bigger squeezing amplitude causes the larger squeezing ratio, furthermore, the squeezing oscillating period is smaller show as Fig.~\ref{fig:osscilation squeezing different R}.

\section{Conclusion}
\label{SectV}

We have systematically investigated the nonlinearity and related phenomenon of the librational mode of an optically levitated ellipsoidal nanoparticle.  By utilizing quantization method, we have found the nonlinearity of librational mode is independent of the frequency and power of laser, and it is only inversely proportional with the rotational inertia of particle.  The nonlinearity of the librational mode is at least two orders larger than the center-of-mass mode of ellipsoidal nanoparticle.
In order to generate the squeezed states for librational mode, we bring a drive to it and find different driving amplitude and frequency can stimulate different steady state. Red-detuning driving induces bistability, but the system is always stable when the driving is blue-detuning. In the stable region, by properly tuning the driving detuning and the strength, we can squeeze the variance of the angle operator in the certain direction, which is useful for the precise measurement of the angle and the angular momentum. In future, we plan to study how to induce the strong
coupling between the librational and the translational modes by properly external driving  \cite{Liu17}. It would be useful for quantum information processing and sympathetic cooling.

\section{Acknowlegement} 
We thank Tongcang Li and Lei-Ming Zhou for helpful discussions.
This work is supported by the NSFC grants No.11374032, 61435007, 11374032, the Joint Fund of the Ministry of Education of China (6141A02011604) and NSAF U1530401.


\end{document}